\documentclass[twocolumn,showpacs,amsmath,amssymb]{revtex4}
\usepackage{psfig}
\usepackage{graphicx}

\begin{document}

\title{\large \bf
Novel two-loop SUSY effects on CP asymmetry in $B\to \phi K_{s}$ }
\date{\today}
\author{\bf  C.H.~Chen$^{a}$\footnote{Email:
physchen@mail.ncku.edu.tw} and C.Q.~Geng$^{b}$\footnote{Email:
geng@phys.nthu.edu.tw} }

\affiliation{$^{a}$Department of Physics, National Cheng-Kung
University, Tainan 701, Taiwan \\
$^{b}$Department of Physics, National Tsing-Hua University,
Hsinchu 300, Taiwan}

\begin{abstract}
Inspired by the exotic measurements on the CP asymmetry in $B\to
\phi K_{s}$, we study a new diagram in supersymmetric models which
can make the difference $\sin2\phi^{eff}_{1}(J/\Psi
K_{s})-\sin2\phi^{eff}_{1}(\phi K_{s})$ to be $20-50\%$ after
satisfying the constraint from $b\to s \gamma$. We also find that
the direct CP asymmetry of $b\to s \gamma$ could be $\sim 10\%$
and testable at $B$ factories.
\end{abstract}

\pacs{ 11.30.Er, 12.60.Jv, 13.25.Hw}

\maketitle

While enjoying the large CP asymmetry (CPA) in the decay of $B\to
J/\Psi K_{s}$ observed by Belle \cite{BelleCP} and Babar
\cite{BabarCP} at the precision level, the recent data on $B\to
\pi^{+} \pi^{-}$ \cite{Belle-80} and $B\to \phi K_{s}$
\cite{Belle-66,Babar-004} have stimulated theorists to think more
about other possible CP violating phases, beside the
Kobayashi-Maskawa (KM) \cite{KM} phase in the standard model (SM).

It is known that with the Wolfenstein parametrization
\cite{Wolfenstein}, the tree and penguin diagrams have the same CP
phase for the inclusive processes of $b\to s \bar{c} c$ and $b\to
s\bar{s} s$. Thus, the time-dependent CPA,
proportional to
$\bar{\Gamma}(\bar{B}\to f_{CP})-\Gamma(B\to f_{CP})$ with
$f_{CP}$ being the final state and having a definite CP property,
arises from the $B-\bar{B}$ oscillation dictated by box diagrams,
in which the source of the CP phase is from $V_{td}=
|V_{td}|e^{-i\phi_{1}}$. For the channel of $f_{CP}=J/\Psi K_{s}$,
the CPA is related to $\sin2\phi_{1}$ and the mixing-induced CP
violation. If there is only the KM phase involved in the
low-energy, the pure penguin process of $B\to \phi K_{s}$  has
approximately the same value of $\sin2\phi_{1}$ as that in the
decay of $B\to J/\Psi K_{s}$, $i.e.$,
 $\Delta S_{\phi_{1}}=\sin2\phi_{1}(J/\Psi K_{s})
-\sin2\phi_{1}(\phi K_{s})
\simeq 0$ \cite{GW}.

It is usually believed that new physics could go into
low-energy phenomena through loop diagrams, in which  new
particles appearing in the loops are integrated out and the
remaining effective couplings are as functions of
their masses and  couplings to the conventional particles.
Since the transition of $b\to s\bar s s$ is a pure
quantum loop effect, one can recognize immediately that $B\to
\phi K_{s}$ is a good candidate to probe new physics.
 Furthermore, although the tree-level contributions
in $b\to s \bar{c} c$ are over a factor of 5 larger than those of
penguin diagrams in the SM \cite{Cheng}, the penguin-type diagrams
induced by new physics could be enhanced, which will clearly
affect the decay of $B\to J/\Psi K_{s}$, especially on its direct
CPA.

To understand the Belle's result of the $3.5\sigma$ difference on
$\sin2\phi_{1}$ between $J/\Psi K_{s}$ and $\phi K_{s}$ modes
\cite{Belle-66}, various theoretical models such as those with
supersymmetry (SUSY) \cite{KK,Dutta,Hou,Datta} and left-right
symmetry \cite{Raidal}
 have been investigated. In addition,
the authors of Refs. \cite{KK,Dutta} have  tried to solve the
problem of unexpected large branching ratios (BRs) in $B\to
\eta^{\prime} K$ decays. However, we would like to address some
problems on these attempts
 as follows:\\
(a) {\it  Direct CP violation on $B\to J/\Psi K_{s}$}: \\
We emphasize that Belle and Babar not only measure an accurate
mixing-induced CPA, but also indicate no direct CPA in $B\to
J/\Psi K_{s}$, up to the percentage level. Those new SM-like
effective four-fermion interactions for
$b\to s\bar{s} s$ will inevitably contribute to $b\to s \bar{c} c$.
 It is also known that there exist large strong phases
in the production of charmed mesons (including charmonium states)
\cite{Chen-PRD,KKLLS}. Therefore, to enhance the BRs of $B\to
\eta^{\prime} K_{s}$ with large CP violating effects will make the direct
CPA in $B\to J/\Psi K_{s}$ to be over the current experimental
limits. \\
(b) {\it
BRs of $B\to \eta K$ and $B\to
\eta^{(\prime)} K^*$}:\\
 We note that the problems for the
production of $\eta^{\prime}$ in $B$ decays depend on not only $B\to
\eta^{\prime} K$, but also $B\to \eta K$ and $B\to
\eta^{(\prime)} K^*$.
 From the data at Babar,  we
have that BR$(B\to \eta K^0 )=(2.9 \pm 1.0 \pm 0.2)10^{-6}$
\cite{Babar-PLB92}, BR$(B\to \eta K^{*0})= (18.6 \pm{2.3}\pm
1.2)10^{-6}$, and BR$(B\to \eta^{\prime} K^{*0})< 7.6 \times
10^{-6}$ \cite{Babar-conf}. By using the perturbative QCD approach
\cite{ES}, we find that the estimating BRs of $B\to \eta K^0$ and
$\eta^{\prime} K^{*0}$  are over the current experimental values,
whereas it is lower for $B\to \eta K^{*0}$.

\begin{figure}[htbp]
\includegraphics*[width=2.in]{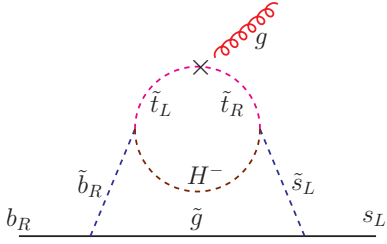}\caption{
Two-loop diagram by the $\tilde{b}_{R}-\tilde{s}_{L}$ flavor changing
effect and the chromodipole operator.}
 \label{twoloop}
\end{figure}

It is clear that to resolve the problems we need more knowledge on
$\eta^{(\prime)}$ mesons as well as their relevant physics. On the
other hand, we may bypass these problems by concentrating on new
physics effects which are insensitive to hadronic uncertainties.
In this paper, we will introduce a two-loop diagram illustrated in
Fig. \ref{twoloop}, in the framework of SUSY models, resulting
from dipole operators. 
In contrast with other mechanisms, such as those discussed in Refs.
\cite {KK,Hou} in which the relevant off-diagonal terms of
squark-mass matrices directly involve flavor changing neutral current that couples to gluino, our two-loop effect shows how
to generate the flavor changing processes
naturally in the SUSY models.
We will illustrate that the diagram not only contributes a sizable value
for the difference of $\sin2\phi_{1}$ between $J/\Psi K_{s}$ and
$\phi K_{s}$ channels, but also satisfies the experimental
constraints such as those from the $b\to s \gamma$ decay and the
neutron electric dipole moment (NEDM).
%
Since the  diagram involves the couplings of the charged Higgs to
squarks, we first discuss the relevant couplings in SUSY models,
given by \cite{Higgs}
\begin{eqnarray}
{\cal L}_{H \tilde{f}
\tilde{f}}&=&-(2\sqrt{2}G_{F})^{1/2}\left(\tilde{V}_{tb}
{\tilde{A}^{b}_{L} \tilde{t}^*_{L}  \tilde{b}_{R}}\right. \nonumber\\
&& \left. + \tilde{V}^{*}_{ts} \tilde{A}^{t}_{R}\tilde{t}^{*}_{R}
\tilde{s}_{L}\right)H^{+} +h.c.,
\end{eqnarray}
where $\tilde{A}^{b}_{L}=m_{b}(A^{*}_{b}\tan\beta -\mu)$ and
$\tilde{A}^{t}_{R}=m_{t}(A_{t}\cot\beta -\mu^{*})$. Here, the
definition of the angle $\beta$ is followed by
$\tan\beta=v_{u}/v_{d}$ with $v_{u}$ and $v_{d}$ being the vacuum
expectation values (VEVs) of Higgs fields $\Phi^u$ and $\Phi^d$
responsible for the masses of upper and down type quarks,
respectively, and $\mu$ is the mixing effects of $\Phi^{u,d}$. For
a large $\tan\beta$ case, $\tilde{A}^{b}_{L}$ and
$\tilde{A}^{t}_{R}$ can be simplified as $\tilde{A}^{b}_{L}\approx
m_{b}A_{b} \tan\beta$ and $\tilde{A}^{t}_{R}\approx -m_{t}
\mu^{*}$. Note that we have neglected the contribution of
$\tilde{s}_{R}$ because the corresponding  coupling is associated
with the strange-quark mass. Moreover, in order to suppress
one-loop contributions, we assume that the flavor mixing effects
on the down-type squark mass matrix are small.

We remark that both flavor changing and chirality flipping are
involved  in Fig. \ref{twoloop}, in which the charged Higgs is
used to change the flavor and the mixing of $\tilde{t}_{L}$ and
$\tilde{t}_{R}$ to govern the chirality flipping, representing by the
cross in the figure. Explicitly, as usual, the
mixing terms are described by  \cite{BurasB520}
\begin{eqnarray}
\left({\bf m}^2_{\tilde{U}}\right)_{LR}&=&\left({\bf
M}^2_{\tilde{U}}\right)_{LR}-\mu \cot\beta {\bf m}_{U},
\end{eqnarray}
where $({\bf M}^2_{\tilde{U}})_{LR}$ represent the trilinear soft
breaking effects. For simplicity, we have adopted the so-called
super-CKM basis, where quarks are in the mass eigenstates so that
${\bf m}_{U}$ is the diagonal upper-type quark mass matrix
\cite{BurasB520}. To overcome the NEDM constraint, it has been
proposed \cite{ABKL} to use hermitian Yukawa and $A$ matrices. The
construction of a hermitian Yukawa matrix can be implemented based
on some symmetries, such as the horizontal $SU(3)_{H}$ \cite{MY}
and left-right  \cite{BDM} symmetries. As a result, the CP phases
of $O(1)$ can exist naturally even with the NEDM contributions.
Moreover, it implies that the CP asymmetries in hyperon decays
could reach the value of $O(10^{-4})$ \cite{Chen}, which is
testable in the experiment E871 at Fermilab \cite{White}.
However, in the class of models proposed in Ref. \cite{ABKL}, the
$\mu$ parameter is real which is not favored in our following
discussions. To avoid this shortcoming, we address the NEDM
constraint by imposing the Yukawa and $A$ matrices to be hermitian
and the squark mass of the first generation to be ${\cal O}(10)$
TeV. Hence, the $\mu$ parameter is regarded as a complex value in
our approach. Due to the hermitian property, a special relation is
obtained as $\left( \delta _{kl}^{U}\right) _{LR}= \left( \delta
_{kl}^{U}\right) _{RL}  \label{mi}$ 
with $(\delta _{kl}^{U})_{LR}\equiv ({\bf M}^2_{\tilde{U}kl})_{LR}
/\tilde{m}^{2}=(V^{U\dagger
}A^{U\dagger}v_{u}V^{U})_{kl}/\tilde{m}^{2}$, where $A^{U\dagger
}= A^{U}$, $V^{U}$ is the mixing matrix for diagonalizing the mass
matrix of upper-type quarks and $\tilde{m}$ is the average squark
mass in the super-KM basis.
In general, the trilinear SUSY soft breaking $A^{Q}$ terms are not
diagonal matrices. However, due to the relation of
$A^{Q}_{ij}=(Y^{Q}\hat{A}^{Q})_{ij}$ with $Y^{Q}\ (\hat{A}^{Q})$
being Yukawa  (A-parameter) matrices and the small effect of
renormalization group, dominant effects of $A^{Q}$ are still from
the diagonal elements \cite{BV}
 if we take $\hat{A}^{Q}$
to be universal and diagonal at the grand unified scale. We use
$A_{Q}=A^{Q}_{ii}$ to simplify our estimations. Therefore, the
contribution in Fig. \ref{twoloop} is proportional to
$m_{b}m_{t}\mu^{*} A_{b}\tan\beta (\delta^{t}_{33})_{LR}$. Since
$A^{U}$ is hermitian, $A^{t}(\delta^{t}_{33})_{LR}$ can be
regarded as real values. Hence, in our mechanism, the CP violating
source is focused on the complex $\mu$ term. We note that by
adopting a large $\tan\beta$, the $\mu$-dependent effect is from
the vertex of the charged Higgs coupling to squarks. For
convenience, we write the relationship between weak  and physical
eigenstates for the mixing of $\tilde{t}_{L}$ and $\tilde{t}_{R}$
as 
\begin{eqnarray}
\left(\begin{array}{c}
  \tilde{t}_{L} \\
  \tilde{t}_{R} \\
\end{array} \right) =
\left( \begin{array}{cc}
  \cos\theta_{t} & \sin\theta_{t} \\
  -\sin\theta_{t} & \cos\theta_{t} \\
\end{array} \right)
\left(\begin{array}{c}
  \tilde{t}_{1} \\
  \tilde{t}_{2} \\
\end{array} \right).
\end{eqnarray}
To study Fig. \ref{twoloop}, we start with the effective
interactions for quark-gluino-squark, given by \cite{HK}
\begin{eqnarray}
{\cal
L}_{\tilde{g}\tilde{q}q}= -\sqrt{2}g_{s}( \bar{ s} P_R
\tilde{g}^{a} T^{a}
 { \tilde{s}}_{L} -\bar{ b} P_L
\tilde{g}^{a} T^{a}  { \tilde{b}}_{R} ) +h.c.,
\end{eqnarray}
 where the
flavor mixings for squarks have been neglected.

It is interesting to note that if we use the photon instead of the gluon
and include the emission of the photon at the charged Higgs, we find that
the same mechanism could also contribute to $b\to s \gamma$.
Therefore,
sizable values for both $\Delta S_{\phi_1}$ and the rate CPA in $B\to
X_{s}\gamma$
can definitely provide a hint for
new physics.
The effective
operators for $b\to s \gamma (g)$ are given by
\begin{eqnarray}
{\cal L} &=&\frac{G_{F}}{\sqrt{2}}V_{ts}^{*}V_{tb}\left(
C_{7\gamma }\left(
\mu \right) {\cal O}_{7\gamma }+C_{8g}\left( \mu \right) {\cal O}%
_{8g}\right) ,\label{eq:op}
\end{eqnarray}
where ${\cal O}_{7\gamma } =m_{b}e/(8\pi ^{2})\bar{s}\sigma _{\mu
\nu }F^{\mu \nu }(1+\gamma_5)b$, ${\cal O}_{8g}
=m_{b}g_{s}/(8\pi ^{2})\bar{s}\sigma _{\mu \nu }T^{a}G^{a\mu \nu
}(1+\gamma_5)b$,
\begin{widetext}
\begin{eqnarray*}
C_{7\gamma }&=& - \cos\theta_{t} \sin\theta_{t}
 \frac{\tilde{V}_{ts}^{*}\tilde{V}_{tb}}{%
V_{ts}^{*}V_{tb}}\frac{\alpha _{s}(m_{b})}{8\pi }
\frac{m_{t}}{m_{\tilde{g}}} \frac{A_{b}\tan\beta\,
\mu^*}{m^{2}_{\tilde{g}}} P_{U}^{\gamma }I\left(
\frac{m_{\tilde{t}}^{2}}{m_{\tilde{g}}^{2}},\frac{m_{\tilde{b}}^{2}}{
m_{\tilde{g}}^{2}},\frac{m_{H}^{2}}{m_{\tilde{g}}^{2}}\right) ,\\
I\left( \frac{m_{\tilde{q}_{1}}^{2}}{m_{\tilde{g}}^{2}},\frac{m_{\tilde{q}%
_{2}}^{2}}{m_{\tilde{g}}^{2}},\frac{m_{H}^{2}}{m_{\tilde{g}}^{2}}\right)
&=&m_{\tilde{g}}^{4}\int_{0}^{1}dx\int_{0}^{\infty
}dQ^{2}\frac{x\left(
1-x\right) Q^{2}}{\left( Q^{2}+m_{\tilde{g}}^{2}\right) \left( Q^{2}+m_{%
\tilde{q}_{2}}^{2}\right) ^{2}\left( m_{\tilde{q}_{1}}^{2}\left(
1-x\right) +m_{H}^{2}x+Q^{2}x\left( 1-x\right) \right) } ,
\end{eqnarray*}
\end{widetext}
$C_{8g} =  C_{7\gamma }/(2N_{c}P^{\gamma}_{U})$ and
$P^{\gamma}_{U}=C_{F}(Q_{U}-1)$ with $C_{F}=4/3$ and $Q_{U}=2/3$
being the color factor and the charge of the upper-type squark, respectively.
Clearly, we
obtain the unique property that the effects of electric and
magnetic dipole moments are directly related to those of
chromoelectric and chromomagnetic dipole moments, respectively.
 Before we
proceed further, we have to examine whether the two-loop effects
are of interest. Explicitly, we would like to check whether the
value of $C_{7\gamma}$ is larger or smaller than experimental
constraint $0.3<|C^{\rm eff}_{7\gamma}|< 0.34$ \cite{btosgamma}.
For an illustration, we set the values of parameters, by
satisfying the constraints from the NEDM \cite{NPB644}, as
follows:
 $\tan\beta\sim m_t/m_b$,
$\sin\theta_{t} \cos\theta_{t} \sim 0.2$,
$\tilde{V}_{ts}^{*}\tilde{V}_{tb}/ V_{ts}^{*}V_{tb}\sim O(1)$,
$A_{b}/m_{\tilde{g}}\sim \mu/m_{\tilde{g}}\sim 4$,
$Arg(\mu)=\pi/2$, $m_{H}\sim 150$ GeV, $m_{\tilde{g}}\sim 1$ TeV,
$m_{\tilde{t}}\sim 200$ GeV and $m_{\tilde{b}}=m_{\tilde{s}}\sim
500 $ GeV, and we have $|C_{7\gamma}|\sim 0.8$. If we take
$\tan\beta\sim 50$, $\sin\theta_{t} \cos\theta_{t} \sim 0.35$,
$A_{b}/m_{\tilde{g}}\sim \mu/m_{\tilde{g}}\sim O(1)$ and the
remains to be the same as the above choices, we obtain
$|C_{7\gamma}|\sim 0.14$.
Furthermore, by using b-quark and sbottom instead of s-quark and
its squark, the similar two-loop diagram could contribute to the EDM
of s-quark. It is known that the current limit of the s-quark
chromo EDM is $|ed^{C}_{s}|_{expt}<5.8\times 10^{-25}\ e\;cm$
\cite{SEDM}. We now examine the contribution to $d^C_s$
in our mechanism.
 By using Eq. (\ref{eq:op}) and assuming
$m_{\tilde{s}}=m_{\tilde{b}}$ and $A_s=A_b$, we obtain
  \begin{eqnarray}
   |d^{C}_{s}|\sim\sqrt{2}G_F V^{2}_{ts} \frac{m_s}{8\pi^2}|Im
   C_{8g}|=\frac{G_F V^{2}_{ts}}{\sqrt{2}C_F} \frac{m_s}{8\pi^2}Im
   |C_{7\gamma}|.
  \end{eqnarray}
Numerically, we get $|ed^{C}_{s}|\sim 2.5 \times
10^{-25}|C_{7\gamma}|\ e\;cm$, which is below $|ed^{C}_{s}|_{expt}$
if $|C_{7\gamma}|\leq 1$.
Clearly, in our mechanism it is inevitable to utilize the large
$\tan\beta$, $A_{b}/m_{\tilde{g}}$ and $ \mu/m_{\tilde{g}}$ scheme
and, therefore,  the most
strict constraint is the BR of $B\to X_{s}\gamma$.

In order to discuss the mixing-induced CP problem in $B\to \phi
K_{s}$, we write the relevant definition of the time-dependent CPA
as
\begin{eqnarray}
A_{CP}&=&{BR(\bar{B}\to \phi K_{s})- BR(B\to \phi K_{s} )\over
BR(\bar{B}\to \phi K_{s})+ BR(B\to \phi K_{s})}, \nonumber \\
&=& C_{\phi K_{s}}\cos\Delta m_{B} t+S_{\phi K_{s}}\sin\Delta
m_{B} t, \nonumber \\
&=&{|\lambda|^{2}-1 \over |\lambda|^2 +1}\cos\Delta m_{B} t-{2{\rm
Im}\lambda \over |\lambda|^2 +1}\sin\Delta m_{B} t, \label{ACP}
\end{eqnarray}
where $\lambda=e^{-i2\phi^{eff}_{1}(\phi K_{s})} A(\bar{B}\to \phi
K_{s})/A(B\to \phi K_{s})$ and $A(B\to f_{CP})$ is the decay
amplitude. Since the dipole operators contributing to
the nonleptonic decays belong to next-to-leading order in
$\alpha_{s}$, we can safely neglect the contributions to the decay
amplitude of $B\to J/\Psi K_{s}$. For displaying the other SUSY
effects on the $B-\bar{B}$ mixing, we use  $\phi^{eff}_{1}$
instead of $\phi_{1}$. Hence,  $\phi^{eff}_{1}$ is still
determined by $B\to J/\Psi K_{s}$, exclusively. For
estimating the hadronic matrix element of $B\to \phi K$, we use
the naive factorization, given by
\begin{eqnarray}
\langle \phi K| {\cal O}_{8g} |\bar{B},p_{B}\rangle &\approx &
-\frac{2\alpha_{s}}{9\pi}\frac{m^{2}_{b}}{q^2} f_{\phi}
m_{\phi}F^{BK}(0)\epsilon^*\cdot p_{B},
\end{eqnarray}
where $F^{BK}(0)$ is the transition form factor of $B\to K$ at
$Q^{2}=0$, $q^{2}$ is the squared momentum of the virtual gluon,
$\epsilon$, $f_{\phi}$ and $m_{\phi}$ correspond to the
polarization vector, decay constant and the mass of $\phi$,
respectively. 
The dominant contribution of factorization assumption is confirmed
by the PQCD approach \cite{MS} in which $q^2$ is related to the
momentum fractions of quarks and convolutes with wave functions.
We note that although ${\cal O}_{7\gamma}$ can also contribute to
the decay of $B\to\phi K_{s}$, since the coupling is
electromagnetic interaction and much smaller than that of strong
interaction, we neglect its contribution. Accordingly, the decay
amplitude for $B\to \phi K_0$ is written as
\begin{eqnarray}
A(\bar{B}\to \phi K_{0})&=&\frac{G_{F}}{\sqrt{2}}V^{*}_{ts}V_{tb}
\left( \sum_{i=3}^{5}a_{i} -\frac{2\alpha_{s}}{9\pi}
\frac{m^{2}_{b}}{q^2}
C_{8g}\right)\nonumber \\
&&\times f_{\phi} m_{\phi}F^{BK}(0)\epsilon^*\cdot p_{B},
\end{eqnarray}
where $a_{i}$, defined in Ref.
\cite{CKL},  stand for the effective Wilson coefficients
in the SM, included from electromagnetic penguin diagrams.
 The value of $\sum_{i=3}^{5}a_{i}$ is estimated to be
$-0.045$. The  parameter $\lambda$ in Eq. (\ref{ACP})
for the CPA can
be simplified as $\lambda=e^{-i2\phi^{\rm eff}_{1}(J/\Psi
K_{s})}e^{-i2\phi_{New}}=e^{-i2\phi^{\rm eff}_{1}(\phi K_{s})}$
with
\begin{eqnarray}
\tan\phi_{New}=-\frac{2\alpha_{s}}{9\pi} \frac{m^{2}_{b}}{q^2} {
{\rm Im}C_{8g} \over \sum_{i=3}^{5}a_{i} -\frac{2\alpha_{s}}{9\pi}
\frac{m^{2}_{b}}{q^2} {\rm Re}C_{8g}}. \label{phi1}
\end{eqnarray}

To display the unique character of the two-loop diagram,
we adopt the value
of $C_{7\gamma}$ such that
 $C_{7\gamma}=-C^{SM}_{7\gamma}\pm i
|{\rm Im}C^{eff}_{7\gamma}|$ and the experimental value
$C^{eff}_{7\gamma}=C^{SM}_{7\gamma}+C_{7\gamma}=\pm i |{\rm
Im}C_{7\gamma}|$ instead of scanning the whole parameter space. By
using $C^{SM}_{7\gamma}=-0.30$ and the identity
$C_{8g}=-3C_{7\gamma}/8 $, the CP violating phase from the decay
amplitude is $\tan\phi_{New}=\mp ( 0.18\pm 0.01^{+0.11}_{-0.06})$,
in which the first error is from $|C^{\rm eff}_{7\gamma}|=0.32\pm
0.02$ and the second theoretical error arises from the uncertainty
in $q^2=(3/8\pm 1/8)m^2_{B}$. Since $S_{\phi K_{s}}=\sin2\phi^{\rm
eff}_{1}(\phi K_{s})$, by taking $\sin2\phi^{\rm eff}_{1}(J/\Psi
K_{s})\approx 0.74$ measured by Belle and Babar, we obtain
$S_{\phi K_{s}}=0.46\pm 0.01^{+0.10}_{-0.21} (0.93 \pm
0.01^{+0.06}_{-0.05})$ where the sign of $\phi_{New}$ is chosen to
be negative (positive). Interestingly, the former value is close
to the central value of the Babar's result \cite{Babar-004}.
 Furthermore, we can straightforwardly calculate the
difference of the CPAs to be
\begin{eqnarray}
\Delta_{\phi^{\rm eff}_{1}}&=&\sin2\phi^{eff}_{1}(J/\Psi K_{s})-
\sin2\phi^{eff}_{1}(\phi K_{s}) \nonumber \\
&=& \left\{%
\begin{array}{c}
  0.28\pm 0.01 ^{+0.21}_{-0.10}\ (-), \\
  \\
  -(0.20\pm 0.01 ^{+0.05}_{-0.06})\ (+). \\
\end{array}%
\right. \label{sincp}
\end{eqnarray}
We now consider the
two-loop effects
for the
CPA in $b\to s \gamma$. According to the
formalism shown in Ref. \cite{KN}, the rate
CPA for $b\rightarrow s\gamma $ is given by%
\begin{eqnarray*}
A_{CP}\left( b\rightarrow s\gamma \right)
&\approx &\frac{1}{100\left| C^{\rm eff}_{7\gamma}\right| ^{2}}
\left\{ 1.1{\rm Im}%
C_{2}C^{\rm eff*}_{7\gamma}\right. \nonumber \\&& \left. +9.52{\rm
Im}C^{\rm eff}_{7\gamma}C^{\rm eff*}_{8g} +0.16{\rm Im}C_{2}C^{\rm
eff*}_{8g}\right\},
\end{eqnarray*}
where $C_{2}\approx 1.11$ and $C^{\rm
eff}_{8g}=C^{SM}_{8g}+C_{8g}$. With the same $C_{7\gamma}$ used
above,  we get $A_{CP}(b\to s\gamma)\approx \pm (10.5\pm 0.6)\%$ for
negative and positive signs in ${\rm Im C_{7\gamma}}$,
respectively. Comparing to the recent Babar's limit of $-0.06 < A_{CP}
(b\to s \gamma) < +0.11$ \cite{Babarbsg}, we find that only
the result with negative sign in Eq. (\ref{sincp}) is reliable,
which could be used to resolve the sign ambiguity
 in ${\rm ImC_{7\gamma}}$. Finally, we remark that although our
upper value on the CP asymmetry of $b\to s \gamma$ is a little bit
over the Babar upper bound, the problem can be removed by
relaxing the required condition $C_{7\gamma}=-C^{SM}_{7\gamma}\pm i
|{\rm Im}C^{eff}_{7\gamma}|$ introduced for our simplified analysis.

In summary, we have studied the novel two-loop SUSY effects on the
CPAs of $B\to \phi K_{s}$ and $b\to s\gamma$. We have found that
with large values of $\tan\beta$ and $A_{b}(\mu)/m_{\tilde{g}}$,
the difference of $\sin2\phi^{\rm eff}_{1}$ between $J/\Psi K_{s}$
and $\phi K_{s}$ can have a deviation of $20-50 \%$. The main
theoretical error is due to the uncertainty in $q^2$.
 We have also shown that the two-loop effect can
 give the CPA in $b\to s\gamma$ around $+10\%$.
It is clear that, since the two-loop contributions to the CPAs in
both decay modes can be the dominant ones in the SUSY models,
experimental measurements at $B$ factories on these CPAs can determine the sizes of these novel contributions.

{\it Note added}:  Our two-loop SUSY mechanism has been applied
to the decay of $B_s\to \mu^+\mu^-$ \cite{Baek}. \\

{\bf Acknowledgments}\\

This work is supported in part by the National Science Council of
R.O.C. under Grant No. NSC-91-2112-M-001-053, No.
NSC-92-2112-M-006-026 and No. NSC-92-2112-M-007-025.

\end{document}